\documentclass[physics,article,accept,pdftex,moreauthors]{Definitions/mdpi}
\firstpage{1} 
\pubvolume{1}
\issuenum{1}
\articlenumber{0}
\pubyear{2022}
\copyrightyear{2022}
\datereceived{} 
\dateaccepted{} 
\datepublished{} 
\hreflink{https://doi.org/} % If needed use \linebreak
\Title{Searching for an Enhanced Signal of the Onset of Color Transparency in Baryons with D$(e,e'p)n$ Scattering}
\TitleCitation{Searching for an Enhanced Signal of the Onset of Color Transparency in Baryons with D$(e,e'p)n$ Scattering}
\Author{Shujie Li 
 $^{1}$, Carlos Yero~$^{2}$, Jennifer {Rittenhouse} 
 West~$^{1}$, Clare Bennett~$^{3}$, Wim Cosyn~$^{3,4}$, Douglas Higinbotham~$^{5}$, Misak Sargsian~$^{3}$ and Holly~Szumila-Vance~$^{5,}$*}

\AuthorNames{Shujie Li, Carlos Yero, Jennifer Rittenhouse West, Clare Bennett, Wim Cosyn, Douglas Higinbotham, Misak Sargsian and Holly Szumila-Vance}
\AuthorCitation{Li, S.; Yero, C.; Rittenhouse West, J.; Bennett, C.; Cosyn, W.; Higinbotham, D.; Sargsian, M.; Szumila-Vance, H.}
\address{%
$^{1}$ \quad Lawrence Berkeley National Laboratory, Berkeley, CA 94720, USA; shujieli@lbl.gov (S.L.); jennifer@lbl.gov~(J.R.W.)   \\
$^{2}$ \quad Department of Physics, Old Dominion University, 
 Norfolk, VA 23529, USA;  cyero@jlab.org  \\
$^{3}$ \quad Department of Physics, Florida International University, Miami, FL 33199, USA; cbenn039@fiu.edu (C.B.); wcosyn@fiu.edu (W.C.); sargsian@fiu.edu (M.S.) \\
$^{4}$ \quad Department of Physics and Astronomy, Ghent University, B9000 Ghent, Belgium\\
$^{5}$ \quad Thomas Jefferson National Accelerator Facility, 
 Newport News, VA 23606, USA; doug@jlab.org }

\corres{Correspondence: hszumila@jlab.org}
\abstract{Observation of the onset of color transparency in baryons would provide a new means of studying the nuclear strong force and would be the first clear evidence of baryons transforming into a color-neutral point-like size in the nucleus as predicted by quantum chromodynamics. Recent C$(e,e'p)$ results from electron-scattering did not observe the onset of color transparency (CT) in protons up to 
spacelike four-momentum transfers squared, 
$Q^2=14.2$ GeV$^2$. 
The traditional methods of searching for CT in $(e,e'p)$ scattering use heavy targets favoring kinematics with already initially reduced final state interactions (FSIs) such that any CT effect that further reduces FSIs will be small. The reasoning behind this choice is the difficulty in accounting for all FSIs.  D$(e,e'p)n$, on the other hand, has well-understood FSI contributions from double scattering with a known dependence on the kinematics and can show an increased sensitivity to hadrons in point-like configurations.  Double scattering is the square of the re-scattering amplitude in which the knocked-out nucleon interacts with the spectator nucleon, a process that is suppressed in the presence of point-like configurations and is particularly well-studied for the deuteron. This suppression yields a quadratic sensitivity to CT effects and is strongly dependent on the choice of kinematics. Here, we describe a possible JLab electron-scattering experiment that utilizes these kinematics and explores the potential signal for the onset of CT with enhanced sensitivity as compared to recent experiments.}


\begin{document}

\section{Introduction}
Quantum chromodynamics (QCD) is the fundamental
theory of the strong interaction and the basis for all nuclear phenomena.
However, the poorly understood nonperturbative dynamics of QCD and the 
phenomenon of confinement make the establishment of the link between QCD and 
nuclear dynamics one of most important tasks of modern nuclear physics.  High energy and momentum transfer (i.e., hard) processes involving nuclei are the 
best testing grounds for understanding the link between QCD and nuclear physics.  The advantage of hard processes is that scattering 
that takes place on an individual quark in the nucleon can be clearly identified.  One topic of great interest for hard processes is elastic scattering from a nucleon 
at large spacelike four-momentum transfers squared, 
$Q^2$.
Within the partonic model of Feynman~\cite{Bjorken:1969ja}, this scattering is predicted to be dominated by single-quark, hard, large-angle elastic scattering as the quark momentum fraction approaches $x_B\rightarrow 1$, a process that does not affect the size of the hadron~\cite{PhysRevLett.24.181}.  In QCD, however, the prediction is different, although also dominated by the $x_B\rightarrow 1$ regime.  The phenomenon of color transparency (CT)~\cite{Brodsky:1988xz,Brodsky:1994kf,Dutta:2012ii,Brodsky:2021dze} as predicted by QCD states that for exclusive processes at high momentum transfer, the cross section is dominated by a hadronic state of very small transverse size, effectively selecting it from the initial state configuration containing an infinite number of states of all sizes. Such a small size state (commonly referred to as point-like configuration~(PLC)) 
is also color-neutral when producing final (observable) hadrons.
An expectation that PLCs are color-neutral in hard exclusive processes constitutes the basis for the prediction of color-transparency phenomena, according to which hard 
exclusive processes probed in nuclear medium should observe reduced final state interaction (FSI) for produced hadrons~\cite{Brodsky:1988xz,Mueller:1988mt}.  This prediction is complicated by the fact that PLCs are not eigenstates of the Hamiltonian. They are quantum--mechanical wave packets and therefore will expand as they propagate in the nuclear medium, thereby reducing the expected CT effects~\cite{Farrar:1988me}. To overcome the expansion effects, the experiments require high enough energy such that Lorentz time dilation will delay such expansion.

In this paper, we discuss how to enhance a potential CT signal by looking for the onset of CT using deuterium breakup and measuring changes in the yields for kinematics of known double scattering and screening regimes of FSIs as a function of increasing momentum transfer. We show that the experiment under consideration is highly sensitive to the final state rescattering of a PLC and that the distance of propagation of the PLC can be suppressed by the external kinematics of the recoil nucleon. We describe in detail how our study extends and builds upon the recent proton CT search results at Jefferson Lab (JLab)
 %SL: this is the convention used by many papers. We always spell out Thomas Jefferson National Accelerator Facility in abstract, then say Jefferson Lab and use JLab afterwards. 
~\cite{PhysRevLett.126.082301}, which did not show any sizable CT effects.

\section{Color Transparency for Mesons vs. Baryons}
is analogous to electric charge transparency in electromagnetic processes, for example, as observed in $\pi^0\to e^+e^-\gamma$ decays through the reduced ionization in the medium by $e^+e^-$ pairs at distances close 
to the decay point~\cite{Perkins:1955aa}. Indeed, several experiments investigating the hard exclusive electroproduction of 
$\rho$ and $\pi$ mesons from nuclei observed signatures consistent with color transparency.  Pion production measurements at JLab reported evidence for the onset of CT~\cite{Clasie:2007aa} in the process $e + A \rightarrow e' + \pi^+ + A^{*}$. The results of the pion electroproduction experiment showed that both the energy and $A$ dependence of the nuclear transparency deviate from conventional nuclear physics and are consistent with models that include CT. The results indicate that the energy scale for the onset of CT in mesons is $\sim$1~GeV.  A CLAS experiment studied $\rho$-meson production from nuclei, and the results also indicated an early onset of CT in mesons~\cite{ElFassi:2012nr}. An increase in the transparency with $Q^2$ for both C and Fe was observed. The rise in transparency was found to be consistent with predictions of CT by models~\cite{Frankfurt:2008pz,Gallmeister:2010wn,Cosyn:2013qe}, which had also accounted for the increase in transparency for pion electroproduction. The $\pi$ and $\rho$ electroproduction data set the energy range to be a few GeV for the onset of CT in mesons. 

The uniqueness of QCD, as encoded in the $\rm SU(3)_C$ symmetry obeyed by strong force interactions, also predicts the formation of 
color singlet 3-quark 
($qqq$)
PLCs.  As a result,
in addition to mesons, one expects to observe CT effects also for 
hard exclusive electroproduction of baryons. Thus, an observation of CT 
in the baryonic sector will indicate a unique interplay of  $\rm SU(3)_C$ symmetry and 
nuclear forces probed in the nuclear medium.  The onset of CT is favored to be observed at lower energy in mesons than baryons since only two quarks must come close together, and therefore the quark--antiquark pair is more likely to form a PLC~\cite{Blaettel:1993rd} than a three quark baryon. {The QCD subfield holographic light-front QCD predicts the onset of CT at 14 GeV$^2$ for the proton~\cite{Brodsky:2022bum}.}  The kinematic regime favorable for the observation of CT effects in the baryonic sector 
is as yet unknown.

The experimental observation of the  signal of the onset of CT in baryons remains ambiguous. The first attempt measured large angle $A(p,2p)$ scattering at the Brookhaven National 
 %Lab
Laboratory
(BNL)~\cite{Carroll:1988rp,Mardor:1998zf,Leksanov:2001ui,Aclander:2004zm}. These experiments measured the transparency as the quasi-elastic cross section from the nuclear target to the free $pp$ elastic cross section. The measurements were taken with perpendicular kinematics which, in electron scattering, is generally dominated by nucleon re-scattering~\cite{Barbieri:2004nf,E97-006:2005jlg}. The results of these experiments indicated a rise in the transparency for outgoing proton momenta of 6--9.5~GeV/$c$ 
 consistent with models of CT; $c$ denotes the speed of light.
However, the transparency was observed to decrease at higher momenta between 9.5 
and 14.4~GeV/$c$. This decrease is inconsistent with CT alone.  Possible explanations for the transparency decrease include an elastic energy-dependent cross section due to nuclear filtering from the Landshoff mechanism~\cite{Kundu:1998ti,Ralston:1990jj} or the excitation of charm resonances beyond the charm-production threshold~\cite{PhysRevLett.64.1011}.  The BNL experiments have the added complication that the incident proton suffers a reduction in flux in medium (these are initial state interactions, ISIs) that must be included in any transparency calculation. 

In the $(e,e'p)$ reaction, however, only the final state proton suffers a reduction in flux and needs to be considered in the measurement. The first experiments using an electron beam to measure CT were at SLAC~\cite{Makins:1994mm,ONeill:1994znv} followed by experiments at JLab~\cite{Abbott:1997bc,Garrow:2001di}. In high $Q^2$ quasielastic $(e,e'p)$ scattering from nuclei, the electron scatters from a single proton, which has some associated Fermi motion~\cite{Frullani}. In the plane wave impulse approximation (PWIA), the proton is ejected without final state interactions with the residual $A-1$ nucleons. The measured $A(e,e'p)$ cross-section would deviate from the PWIA prediction in the presence of FSI, where the proton can scatter both elastically and inelastically from the surrounding nucleons as it exits the nucleus. 
Here, the nuclear transparency, $T$, 
of a 
nucleus $A$ 
is defined as (see~\cite{PhysRevLett.126.082301})\\
\begin{equation}
T(A) = \frac{\textrm{measured cross section}}{\textrm{PWIA calculation}}
\end{equation}

In complete CT, the FSIs vanish, and the nuclear transparency plateaus at $T=1$. 
This is in contrast to the conventional picture for the nuclear transparency
in  $A(e,e'p)$ processes in which the $Q^2$ is small.
There, the transparency is less than 1 due to FSIs and other correlations and follows the same energy dependence as the $NN$ cross section, which is relatively constant for momenta above the few GeV range.

None of the earlier $A(e,e'p)$ experiments observed the onset of proton CT. However, the onset of CT depends both on momentum and energy transfers, affecting the ``squeezing'' and ``freezing'' of PLCs~\cite{Frankfurt:1993es}. 
Since $A(e,e'p)$ scattering measurements are carried out at $x_B=1$ kinematics, they have a high Lorentz factor characterized by lower energy transfers compared to momentum transfers. 
It is therefore possible that the JLab 6~GeV era experiments were unable to satisfy the energy dependence of the freezing requirement, even if the momentum transfer was sufficient to create a PLC. The most recent result on this topic was published with the $C(e,e'p)$ data in Hall C at JLab as part of the 12~GeV program~\cite{PhysRevLett.126.082301}. With the beam energy almost doubled as compared with previous JLab measurements, this experiment eliminated the possibility that the increase in transparency observed at BNL was due to the higher incident proton momentum than was attainable during the 6~GeV era.

Results from the recent CT experiment~\cite{PhysRevLett.126.082301} do not indicate a rise in CT up to \linebreak{$Q^2=14.2$~GeV/$c^2$} within a 6\% 
uncertainty. It should be noted that these results searched for the onset of CT in kinematics 
in which the initial momentum of the struck proton was small and the FSI effect was dominated by 
screening effects due to the interference of PWIA and single rescattering amplitudes.  Such kinematics for medium nuclei are characterized by small amplitudes of FSI, and the 
rescattering process takes place over the full radius of the nucleus.
As a result, the transparency $T$ has a  diminished sensitivity to CT effects, both due to 
the possibility that the PLC was not fully formed at the available $Q^2$ and/or that the PLC could undergo a full expansion during 
propagation in the nuclear medium after $ep$ scattering.  Thus the lack of the CT effects in the hard exclusive production of 
the proton from nuclei indicates that the effects are much smaller than 
initially expected, either due to the difficulty in production of small-size PLCs at the 
available $Q^2$ or due to a fast expansion of a 3-quark PLC in nuclear medium.

\section{The Role of D\texorpdfstring{\boldmath$(e,e'p)n$}{(e,e'p)} in CT Studies}
\label{sec:deuteron}
The deuteron is the simplest target nucleus, and its dynamics in collisions are well-described using {generalized 
 %Eikonal
eikonal
approximation (GEA)~\cite{Capel:2019zor,Sargsian:2004tz}.} With a deuteron target, the kinematics in exclusive processes
can be precisely chosen such that the inter-nucleon distances of the struck and spectator nucleon lead to well-controlled FSIs~\cite{Frankfurt:1994kt}. 
In the case of high $Q^2$, with the onset of the CT regime,
these reactions enable observations of the formation of the PLC before expansion and 
control its expansion in the FSI process.

\begin{figure}[H]%[htb]
%\centering
\includegraphics[width=0.97\textwidth]{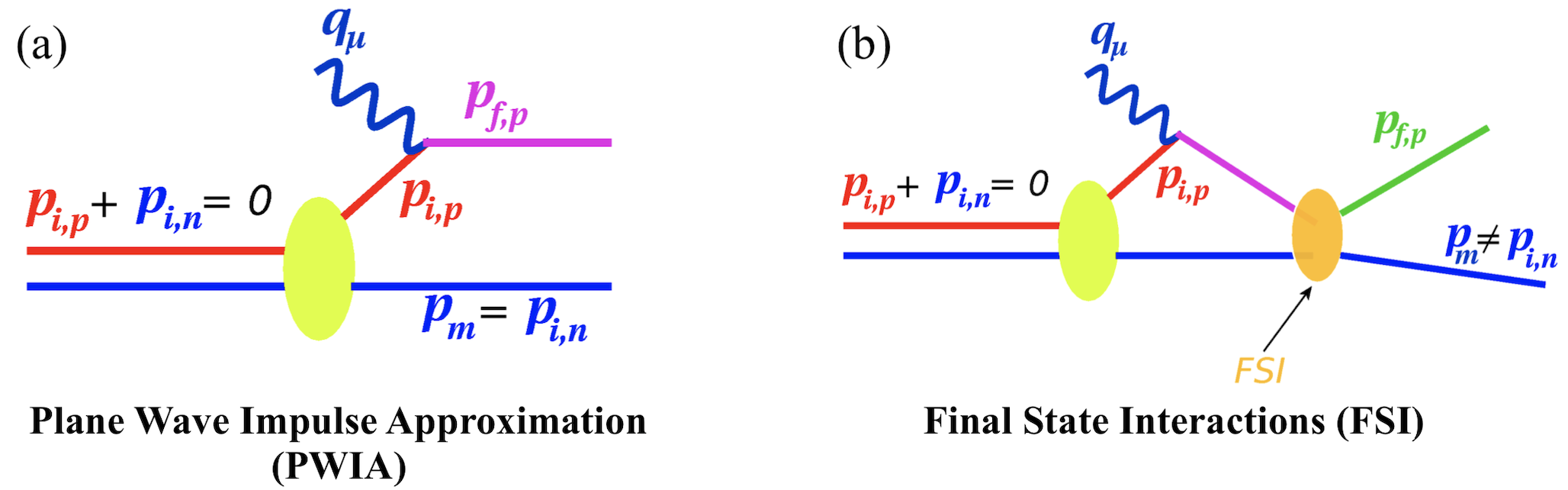}
\caption{\label{fig:feynmanRescatter} Electron scattering interaction probing deuterium (green blob) 
with (\textbf{a}) breakup and (\textbf{b}) re-scattering effects in the initial deuterium rest frame~\cite{Yero:2020urm}.
See text for details.}
\end{figure}

\begin{figure}[H]%[htb!]
%\centering
\includegraphics[width=0.9\textwidth]{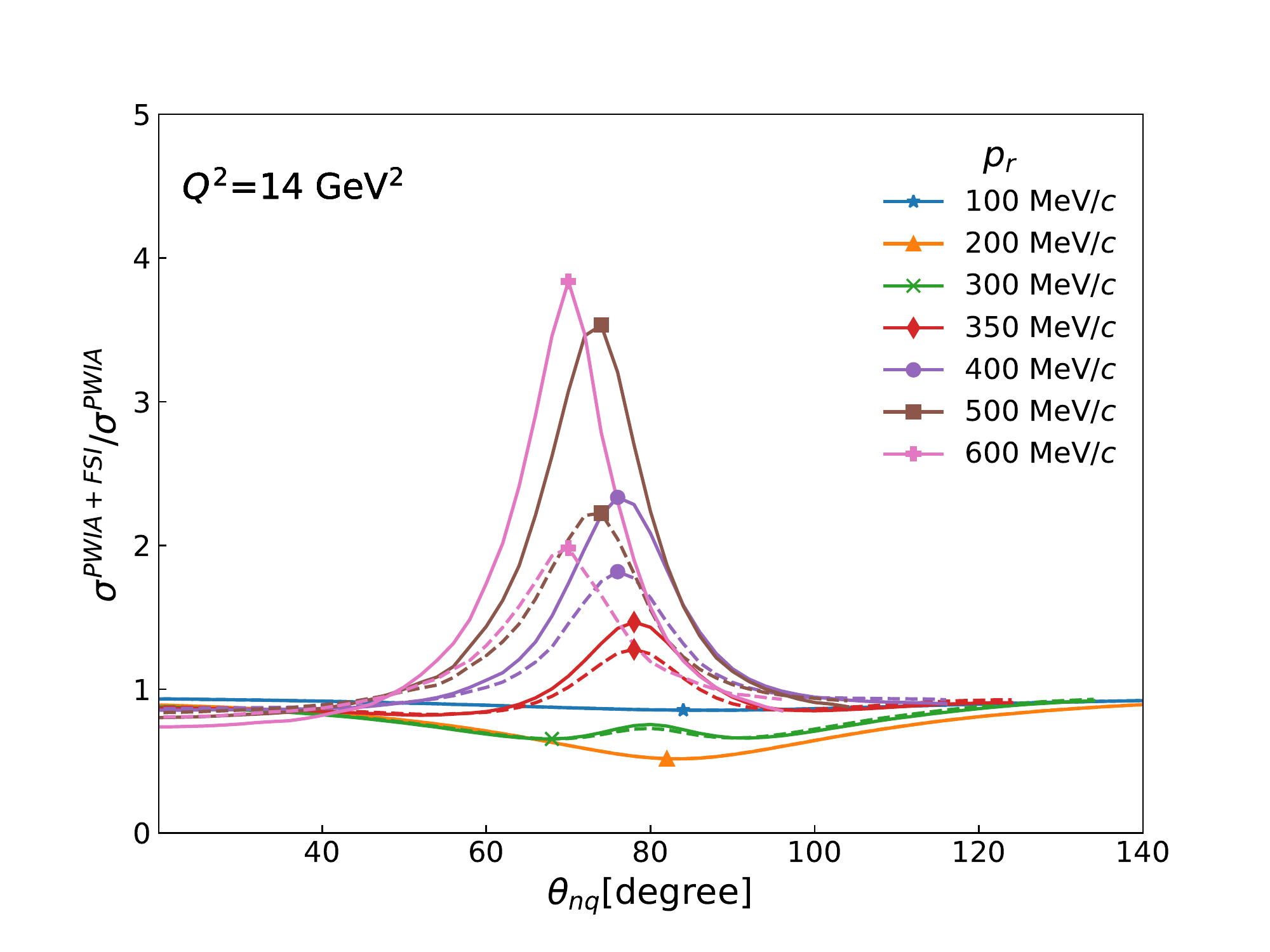}
\caption{\label{fig:rescatterCalc2} The calculated cross-section ratio with respect to the PWIA
 shown for various neutron recoil angles $\theta_{nq}$ (i.e., angle with respect to $\vec{q}$, the 3-momentum of the incoming photon)
at different values of the recoil neutron momenta, $p_r$,  
for $Q^2=14$~(GeV$/c)^2$. Solid curves are calculated with FSI+PWIA~\cite{Sargsian:2009hf}. Dashed curves with the same color and marker are the same calculation, including a 
CT) effect at $\Delta M^2 = m^2_\text{ex}- m^2_h=2$ GeV$^2$, where $m_\mathrm{ex}$, $m_h$ are the masses of the proton in the excited intermediate state and the ground state, respectively.
The AV18~\cite{av18} potential was used in this calculation. This ratio is 
insensitive to 
the choice of potentials up to $p_r=400$ MeV.}
\end{figure}

The proton knockout reaction in deuterium breakup D$(e,e'p)n$ is described by
\begin{equation}
    e+D\rightarrow e'+p+n,
\end{equation}
where the final state proton carries almost all of the momentum transferred to the deuteron and can be identified as a struck nucleon.   {Note that 
the experimental missing momentum $p_m=p_{f,p}-q$ 
in the deuteron is equivalent to the recoil neutron momentum, $p_r$, which is fixed by kinematics. 
Here, $p_{f,p}$ and $q$ are the final momentum of the proton and the momentum of the incoming photon, respectively. 
In kinematics, in which $Q^2\ge 1$~GeV$^2$, two main scattering amplitudes  describe the dynamics of the process (Figure~\ref{fig:feynmanRescatter}). In PWIA, the initial momenta of the struck proton $p_{i,p}=-p_{r}$.
For small $p_{r}$ ($\le$200~MeV/$c$), 
the 
PWIA amplitude dominates over the re-scattering amplitude (as described in Figure~\ref{fig:feynmanRescatter}). With increasing $p_r$, the PWIA amplitude dies out much faster than the re-scattering amplitude, which dies out over a much larger scale of momentum transfer due to the deuteron wave function. }
Therefore, the deuteron cross section's deviation from PWIA is determined mainly by Glauber screening effects ({i.e.,} the interference between PWIA and FSI amplitudes). As a result, the transparency as defined by the ratio of the full cross section to the PWIA cross section is less than unity (see the $100$ and 
$200$~MeV/$c$
curves in Figure~\ref{fig:rescatterCalc2}). This effect is relatively 
small, and thus, deuterium is traditionally not a good target for CT measurements for the case of 
small initial momenta.
 
 %\unskip
%\vspace{-6pt} 

However, if $p_{r}$ is selected to be above 
$300$~MeV/$c$, 
the FSI amplitude (Figure~\ref{fig:feynmanRescatter}b) in near perpendicular kinematics (i.e., the recoil neutron angle is nearly perpendicular to the momentum transfer) significantly exceeds the PWIA 
contribution.  As a result, the transparency as described above will exceed unity (see the $400$, $500$ and 
$600$~MeV/$c$  
curves in Figure~\ref{fig:rescatterCalc2}). 
This demonstrates that the experimental kinematics can easily be tuned such that FSIs dominate the signal. In this situation, the measured cross sections are very sensitive to any change in FSI effects, such as the onset of color transparency.

According to our discussion for perpendicular kinematics, CT will increase the full cross section 
at small recoil neutron momenta 
($p_r\le 200$~MeV/$c$) 
due to the decrease in Glauber screening effects. At the same time, CT will  decrease the full cross section at \mbox{$p_r> 
400$~MeV/$c$} 
due to the decrease in double scattering contribution.
Thus, the experimentally measured D$(e,e'p)n$ cross-section ratio of high to low recoiling neutron momentum at perpendicular kinematics can be used to track the size of FSI effects with 
higher sensitivity than in traditional measurements of CT effects in 
$ee'p$ processes on medium to heavy nuclei. Indeed, in this case, CT makes FSI effects smaller, thereby decreasing the large $p_r$ cross section (numerator) and increasing the small $p_r$ cross section (denominator). As $Q^2$ increases, the onset of CT can be detected through a decrease in such ratios as compared to PWIA calculations. A possible way of performing such a high sensitivity CT measurement with the deuteron at JLab is discussed in Section~\ref{sec:new_proposal}.

\subsection{Recent D\texorpdfstring{$(e,e'p)n$}{(e,e'p)} Studies at High \texorpdfstring{$Q^{2}$}{Q2}}
The first studies of D$(e,e'p)n$ at high $Q^{2}$ (> 1 (GeV/$c)^{2}$) were carried out at JLab experimental Halls A~\cite{HallA:2011gjn} and B~\cite{PhysRevLett.98.262502}. Both experiments observed a strong angular
dependence of the D$(e,e'p)n$ cross section with neutron recoil angle, $\theta_{nq}$, peaking at $\theta_{nq}\sim70^{\circ}$ in agreement with 
the GEA predictions~\cite{Sargsian97}. Due to statistical limitations in Hall B, however, it was necessary to integrate over a wide angular range ($\theta_{nq}$) which resulted in the extracted  deuteron momentum distributions (above $p_{r} \sim$ 300 MeV/$c$) being dominated by FSIs, meson-exchange currents (MECs) and bound-nucleon excitations into a resonance state (isobar currents or ICs). 

The Hall A high-resolution spectrometers, on the other hand, allowed for a measurement of the recoil momentum dependence of the D$(e,e'p)n$ cross sections for fixed bins in $\theta_{nq}$. Different theoretical models (i.e., Paris~\cite{Paris:1981} and 
{charge dependent CD}-Bonn-based~\cite{CDBonn:2001}
calculations) were tested for recoil momenta between $\sim$300 and 600 MeV/$c$. The results from~\cite{HallA:2011gjn} also revealed a kinematic window at $\theta_{nq}\sim$ 35--45$^{\circ}$ where the FSIs are significantly reduced and approximately independent of the recoil nucleon momentum.  An extension of the D$(e,e'p)n$ cross-section measurements beyond $p_r\sim500$ MeV/$c$ is possible and may potentially reveal a transition between nucleonic and quark--gluon degrees of freedom. 

 {Recent studies of deuterium electro-disintegration ~\cite{cyero_2020} carried out in JLab Hall C used the kinematic window ($\theta_{nq}\sim$ 35--45$^{\circ}$) found in ~\cite{HallA:2011gjn} to extend the D$(e,e'p)n$ cross-section measurements to very high recoil momenta ($p_{r}\sim$500--1000 MeV/$c$) and large $Q^{2}$ ($\sim$4--5 (GeV/$c)^{2}$) and 
Bjorken
$x_{\rm Bj} %{bj}
\equiv Q^{2}/2M_{p}\omega>1$, where 
%($M_{p}$, $\omega$) 
$M_{p}$ and  $\omega$
are the proton mass and energy transferred to the nucleus, respectively. At these kinematics, theoretically, MECs and ICs are expected to be suppressed. The suppression of MECs is understood from the fact that the estimated MEC scattering amplitude (i.e., its propagator) is proportional to $(1+Q^{2}/m_\text{meson}^{2})^{-2}(1+Q^{2}/\Lambda^{2})^{-2}$, where $m_\text{meson}\approx $0.71 GeV/$c^{2}$ and $\Lambda^{2}\sim$ 0.8--1 (GeV/$c$)$^{2}$~\cite{Sargsian:2001ax}. Since the IC contributes in kinematics 
above the $\Delta$-isobar threshold, which corresponds to 
 %$x_{bj}<1$, 
$x_{\rm Bj}<1$, 
it is a small correction at 
$x_{\rm Bj}>1$.
Additionally, the 
IC contribution is suppressed dynamically due to the fact that its production amplitude is proportional to the spin--flip electromagnetic amplitude, which falls off with $Q^2$ much faster than elastic form factors~\cite{Stoler:1993yk}.
}
%The contributions from ICs can be suppressed kinematically by selecting 
%$x_{bj}>1$, 
$x_{\rm Bj}>1$,
where one probes the low-energy ($\omega$) part of the deuteron quasielastic peak, which is maximally away from the inelastic resonance production threshold.

The results demonstrated that it is possible to select bins in $\theta_{nq}$ that minimize re-scattering effects and enhance the PWIA, allowing for a direct probe of the least well-known part of the $NN$ interaction at the
sub-Fermi (<1 fm) distance scale, where nucleons begin to significantly overlap.

The extracted momentum distributions from~\cite{cyero_2020} were compared to the theoretical momentum distributions based on either the Paris~\cite{Paris:1981} (calculated by 
J. M. Laget~\cite{LAGET200549}), AV18~\cite{av18} or CD-Bonn~\cite{CDBonn:2001} (calculated by M. Sargsian~\cite{Sargsian:2009hf}) potentials. For recoil 
momenta below 400 MeV/$c$, all theoretical calculations agree with the data; however, at larger recoil momenta, the data were best reproduced by the calculations using the CD-Bonn~\cite{CDBonn:2001} potential up to recoil momenta~700 MeV/$c$. Beyond this momentum range, current theoretical calculations start to break down and are unable to describe data. 

Nevertheless, the progress made in exploring the short-range part of the NN interaction
using deuterium demonstrated that, unlike in heavier nuclei, 
FSIs are well-understood from both an experimental and theoretical perspective. It was also 
demonstrated, particularly in~\cite{cyero_2020}, that it is possible to select specific kinematics of $\theta_{nq}$ to either enhance or suppress the effects of FSI. This last point is crucial
in CT studies using the deuteron, as 
described in  Section~\ref{sec:ct-in-deut}.

\subsection{PLC Expansion Effects}
 {An important aspect of CT is that the PLC will expand as it moves through the nucleus because it is not an eigenstate of the Hamiltonian.  The expansion time,  
$t$ (or equivalently, coherence length, $l_c$, 
over which the hadron travels, evolving from a PLC to the normal size), 
is unknown, but there are several estimation methods in the literature.  A widely used estimate gives $t \approx (E_h / m_h) t_{0}$, where $E_h$ is the 
laboratory (lab) %lab  
energy, $m_h$ is the mass of the ground state hadron, 
and $t_{0} \approx 1 ~\mathrm{fm}$ is the characteristic hadron rest frame time~\cite{Frankfurt:1992zp}, which shows that an increase in $E_h$ allows a PLC to remain small for a longer time, thus increasing the suppression of FSIs.}

 {For the knocked-out proton in the kinematics of the recent experiment~\cite{PhysRevLett.126.082301}, i.e., for $Q^2$ = 8--14 GeV$^2$, holographic light front QCD (HLFQCD) techniques give a coherence length of $l_c \sim $ 2--3 fm~\cite{Caplow-Munro:2021xwi}. This is more than sufficient to observe CT of the proton in deuterium, but we note that there are a few differences in analysis from the recent HLFQCD 
%work
paper  %of~
 \cite{Brodsky:2022bum}.  The naive parton model in which PLC constituents separate at the speed of light gives a transverse hadron 
size of $x_{t} \sim t \sim(E_h / m_h)^{-1} z$, where $z$ is the longitudinal coordinate in the lab frame and $E_h/m_h$ is the time dilation factor as before. This gives a coherence length estimate of $l_{c} \approx E_h / m_{h} ~\mathrm{fm}$ for the PLC~\cite{Farrar:1988me}. The uncertainty principle gives a maximum coherence length, $l_{c} \sim \frac{1}{\Delta M} \frac{p'}{m_{h}}$, where $\Delta M$ is a characteristic excitation energy and $p'$ is the 3-momentum of the particle in the intermediate excited state~\cite{Sargsian.2003}.  The quantum diffusion model is inspired by perturbative QCD (pQCD) and gives a coherence length $l_{c} \simeq\left\langle 1 /\left(E_{\rm ex}-E_{h}\right)\right\rangle \simeq 2 p'\left\langle 1 /\left(m_{\rm ex}^{2}-m_{h}^{2}\right)\right\rangle$, where $m_\text{ex}$ is the mass of the hadron in the excited (i.e., intermediate) state~\cite{Farrar:1988me}. This can be seen by considering the intermediate state propagator: $l_c = {1\over E_\text{ex}-E_h}$, with $E_h = \sqrt{m_h^2+p^2}$ and $E_\text{ex} = \sqrt{m^2_\text{ex}+p'^2}$. For high-energy kinematics, $p'\gg m_h, m_\text{ex}$, the coherence length becomes 
$l_c\approx {2p'/\Delta M^2}$,
where $\Delta M^2 = m^2_\text{ex}- m^2_h$.  The smaller size of deuterium assists in minimizing the expansion effects by minimizing the PLC travel length prior to collision site exit.
}
\section{Constraints on the Color Diffusion Model % from $^{12}$C$(e,e'p)$}
from \texorpdfstring{\boldmath$^{12}$C$(e,e'p)$}{12C(e,e'p)}}

The flat $Q^2$-dependence of the recent Hall C CT measurement~\cite{PhysRevLett.126.082301} is in agreement with standard Glauber calculations not including any CT effects. Calculations including CT effects using the color diffusion model~\cite{Farrar:1988me,Frankfurt:1994kt} with standard parameters ($\Delta M^2=1~\text{GeV}^2$) overshoot the data by a large amount for the highest $Q^2$ values~\cite{PhysRevLett.126.082301,Cosyn:2022msj}.  It is possible, however, to modify the parameters in the color-diffusion model to see which values would still be compatible with the measured Hall C data.  These values could then be used to determine the optimal choice of kinematics to observe the onset of CT in deuterium (see Section~\ref{sec:new_proposal}; we find perpendicular kinematics to be the best choice).

In Glauber calculations using the quantum diffusion model, the total cross-section parameter entering the Glauber profile function is replaced with a position-dependent one.  
The effective total
 cross section, $\sigma^{\text{eff}}$, 
evolves from a reduced value (reflecting the small-sized color transparent configuration) to its normal 
\highlighting{value, $\sigma^{\text{tot}}$,
along} 
a coherence length $l_c$, where the $z$ coordinate is along the momentum of the ejected particle with $z=0$ at the point of interaction with the probe:
\begin{equation}
\sigma^{\text{tot}} \longrightarrow
{ \sigma^{\text{eff}}(z) } =   
{ \sigma^{\text{tot}}} \biggl\{ \biggl[
 \frac{z}{l_c} +
 \frac{ n^2 \langle k_t \rangle^2}{Q^2} \left( 1- \frac{z}{l_c} 
\right) \biggr]
\theta(l_c-z) + 
\theta(z-l_c) \biggr\} \,  \; .
\label{eq:diffusion}
\end{equation}

Here, $l_c=2p/\Delta M^2$, with $p$ as the momentum of the knocked-out proton.
For the parameters in $\sigma^{\text{eff}}$, a standard assumption is $\Delta M^2=1.0~\text{GeV}^2$,  where $n=3$ reflects the elementary constituents in the proton, and 
 %$\langle k_t=0.35$ GeV/$c \rangle$ 
 \highlighting{$\langle k_t\rangle = 0.35$ ${\rm GeV}/c$} 
 the average transverse momentum of a quark inside a hadron~\cite{Farrar:1988me}.  One can see that the effective cross section is reduced at the interaction point $z=0$ to $\sigma^\text{eff} 
(z=0) = \frac{ n^2\langle k_t\rangle^2 }{Q^2}\sigma^\text{tot}$ and then rises linearly over the coherence length $l_c$ to its standard value $\sigma^\text{tot}$.

In order to map the quantum diffusion model parameter values that would still agree with the recent Hall C data, we perform calculations within the relativistic multiple scattering Glauber approximation (RMSGA)~\cite{Ryckebusch:2003fc,Cosyn:2006vm,Cosyn:2022msj} with modified parameters: we take modified $\Delta M^2$ values, which results in a modified coherence length, 
and introduce a new modification parameter, $\alpha$, such that
\begin{equation}
    \mathcal{R} = \alpha \frac{ n^2\langle k_t \rangle^2}{Q^2}\,,\label{eq: sig-ratio}
\end{equation}
and 
\begin{equation}
 \sigma^{\text{eff}}(z=0) = \mathcal{R}\sigma^\text{tot} = \alpha \frac{ n^2\langle k_t \rangle^2}{Q^2}\sigma^\text{tot}   
\end{equation}
is now the modified cross section at the interaction point.

\begin{figure}[H]%
   % \centering
    \includegraphics[width=0.85\textwidth]{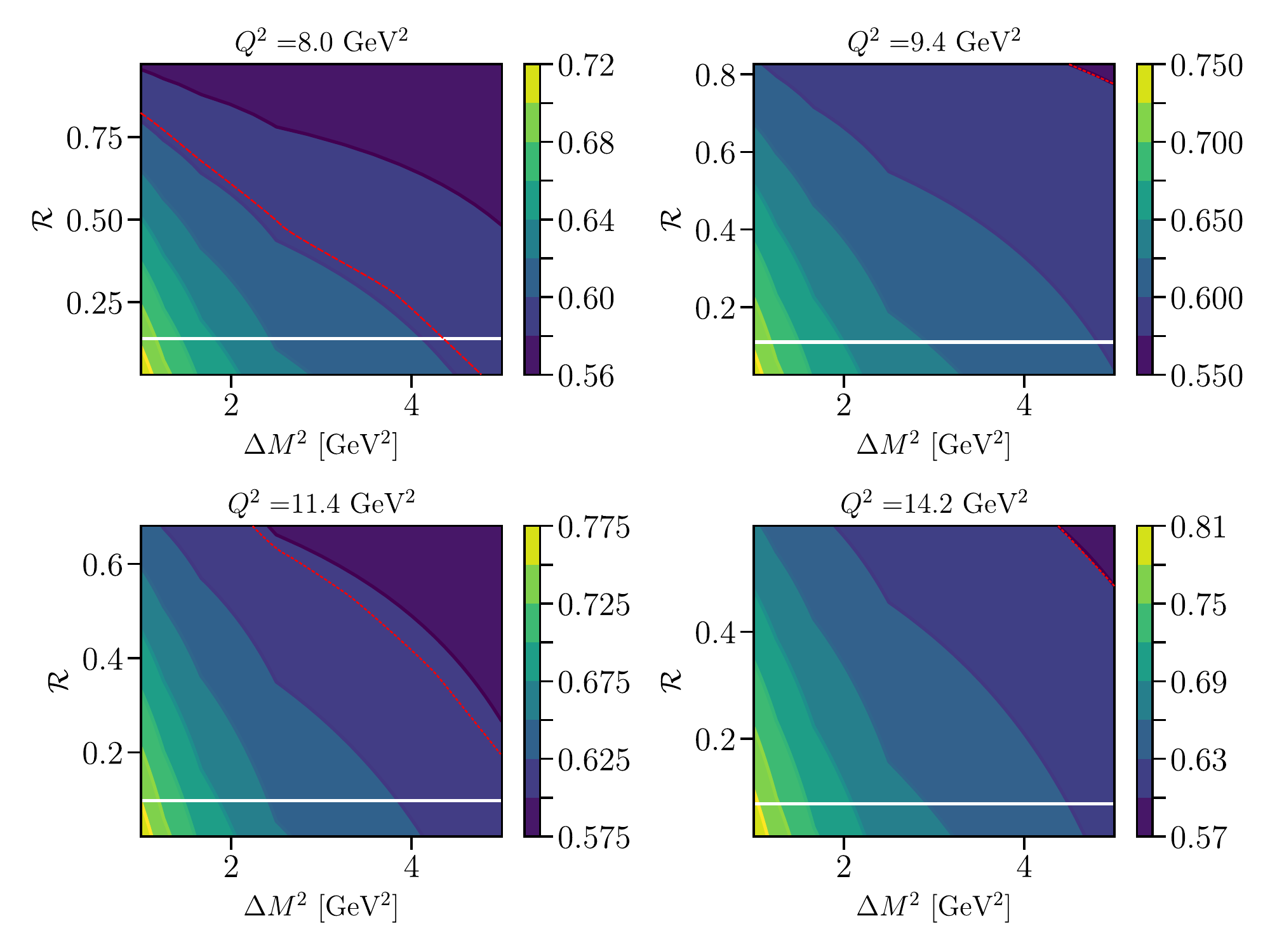} 
    \caption{Transparencies resulting from RMSGA calculations with CT effects included in the quantum diffusion model.  The dependence on the 
$\Delta M^2$ parameter ($x$-axis) and ratio $\mathcal{R} = 
%\frac{\sigma_\text{eff}(z=0)}{\sigma^\text{tot}}$ 
{\sigma^\text{eff}(z=0)}/{\sigma^\text{tot}}$ 
at the interaction point ($y$-axis) is shown.  \textls[-15]{The four panels show the four measured $Q^2$ bins of the recent JLab experiment. Red dashed lines on the plots delineate the region (above the red line) compatible with the recent JLab data~\cite{PhysRevLett.126.082301} at the 
 %1$\sigma$ 
one standard deviation level.  White horizontal lines denote the $\mathcal{R}(\alpha=1)$ standard value. See text for details.
}}
    \label{fig:qdiff}
\end{figure}

Figure~\ref{fig:qdiff} shows which parameter values would still be compatible with the recent Hall C data.  
 %It is clear 
One can see that the highest $Q^2$ data provide the most stringent constraints and that coherence lengths corresponding with $\Delta M^2$ of 3 GeV$^2$ or more are required.

\section{Probing Proton CT in Double-Scattering Kinematics}
\label{sec:ct-in-deut}
\begin{figure}[H]%[htb]
%\centering
\includegraphics[width=0.85\textwidth]{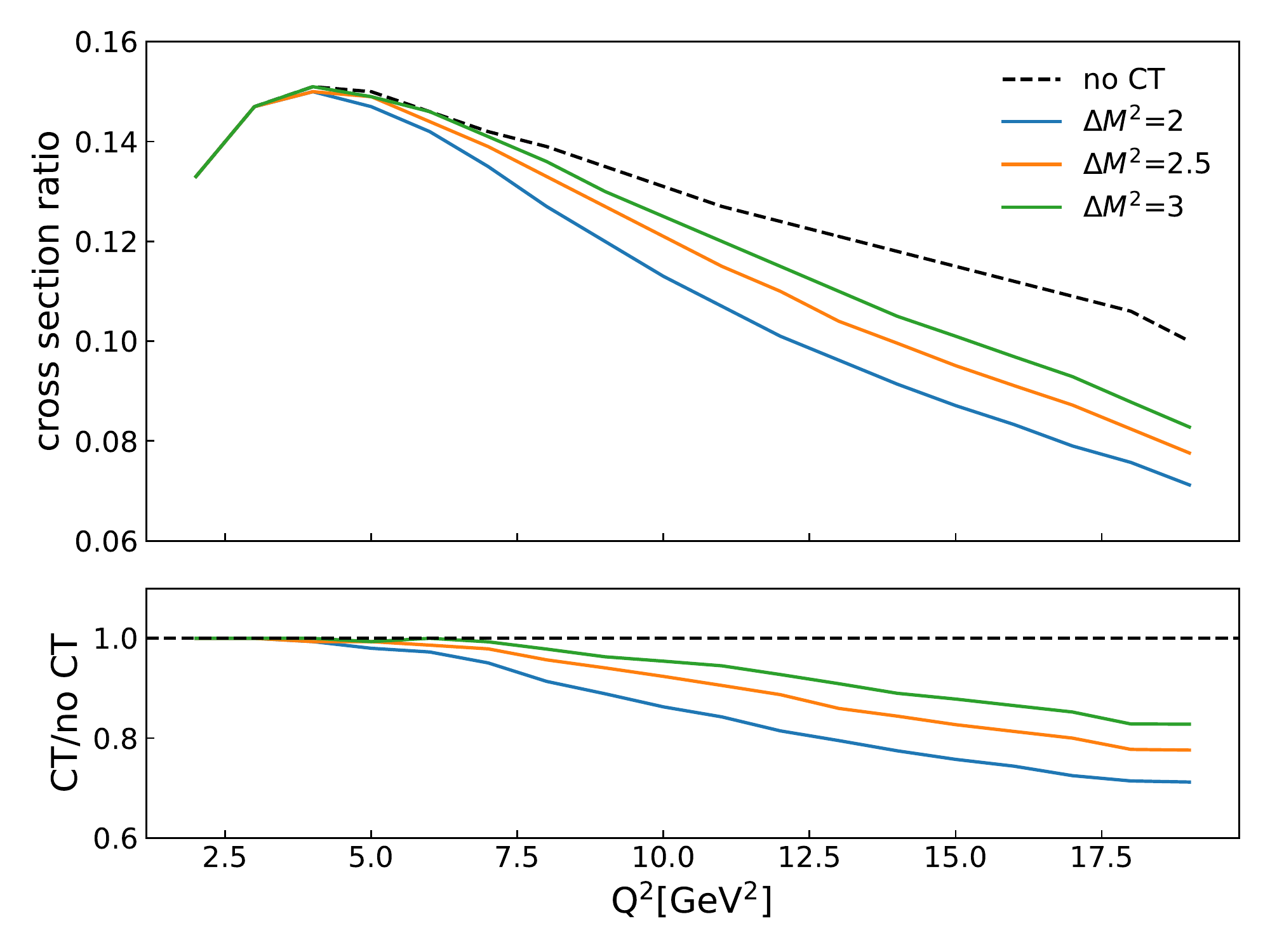}
\caption{\label{fig:rescatteringPrediction} Predictions of cross section ratios, 
$R^{\text{theory}}_D=%\frac
{\sigma(p_r=400~\textrm{MeV/}c)}/{\sigma(p_r=200~\textrm{MeV/}c)}$,  
in the D$(e,e'p)n$ reaction at $\alpha_n \approx 1$ for various hypotheses of $\Delta M^2$~\cite{Farrar:1988me} that are compatible with the recent Hall C results~\cite{PhysRevLett.126.082301}. Here, $\alpha_n$ is the light cone momentum fraction of the nucleon as defined in 
Equation~(\ref{eq:alpha_n}).
{\bf Top:}  
predicted cross-section ratios include the effect of CT with $\Delta M^2$ between 2 and 3 GeV$^2$. 
{\bf Bottom:} 
the change of cross section ratio 
with various $\Delta M^2$ values comparing to "no CT" calculations performed in the generalized eikonal approximation (GEA)  
framework~\cite{Sargsian97,Sargsian:2009hf} using the AV18 
deuterium wave function.}
\end{figure}

The recent proton results \cite{PhysRevLett.126.082301} reinvigorated the physics community and have prompted significant ongoing discussion as well as a re-examination of CT models. It is possible to investigate the effects of CT in the deuteron where the FSI and kinematics are well-understood, as described in Section~\ref{sec:deuteron}.  Double scattering accesses inter-nucleon distances on the order of 1--2~fm. Access to distances of this magnitude enables observations of any PLCs and would help constrain the PLC expansion rate, which may be larger than expected due to the lack of recent experimental observation~\cite{Sargsian.2003}. The larger the momentum of the spectator nucleon, the smaller the inter-nucleon distance, and thus the shorter the distance between the production and re-scattering vertices, leading to higher FSI effects. While the idea to explore CT in re-scattering (perpendicular) kinematics was proposed prior to the 12~GeV upgrade of JLab~\cite{Sargsian.2003}, updated predictions from Sargsian include the cross section from the recent JLab Hall C D$(e,e'p)n$ measurement~\cite{cyero_2020} and implementation of CT effects in the color diffusion model using the new constraint of $\Delta M^2>2$~GeV$^2$ implied from the recent C$(e,e'p)$ CT results. These new predictions indicate that there is parameter space to observe the signal of CT at $Q^2>8$~(GeV$/c)^2$ as shown in Figure~\ref{fig:rescatteringPrediction}.  It is important to note that the new calculations only change $\Delta M^2$ and do not modify the 
ratio
$\mathcal{R}=%\frac
{\sigma^\text{eff}(z=0)} %_\text{eff}(z=0)} 
/{\sigma^\text{tot}}$.
 {Here,} 
the cross section ratio on the vertical axis is
\begin{equation}
R^{\text{theory}}_D(Q^2)=\dfrac{\sigma(p_r=400~\textrm{MeV/}c; Q^2)\downarrow}{\sigma(p_r=200~\textrm{MeV/}c; 
Q^2)\uparrow},\label{eq:rd_theory}
\end{equation}
 the ratio between high ($p_r=400$~MeV$/c$) and low ($p_r=200$~MeV$/c$) recoiling neutron momenta in the D$(e,e'p)n$ reaction at perpendicular 
kinematics, i.e., the neutron recoil angle $\theta_{nq}$ is between 60$^{\circ}$ 
to 90$^{\circ}$ such that FSI effects deviate most from the PWIA calculation as shown in Figure~\ref{fig:rescatterCalc2}. More specifically, in this calculation we require the light cone momentum fraction of the nucleus carried by the recoil nucleon, $\alpha_n \approx 1$, with 
\begin{equation}
    \alpha_n = \frac{(E_n-p_r\text{cos}\theta_{nq})}{m_n}\, ,   \label{eq:alpha_n}
\end{equation}
expressed in the deuterium rest frame in terms of the final state spectator nucleon energy $E_n$, momentum $p_r$, mass $m_n$, and angle 
with the virtual photon $\theta_{nq}$. The high $p_r$ region is dominated by double scattering, the large increase in the cross section 
from PWIA prediction is shown in data from~\cite{PhysRevLett.98.262502}. As $Q^2$ increases, we expect to see an increased probability of 
scattering from a PLC (and hence experiencing less FSI effects) so that the observed cross section decreases at high $p_r$, and increases 
at low $p_r$ as indicated by the down \highlighting{($\downarrow$)} and up \highlighting{($\uparrow$)} 
arrows in 
Equation~(\ref{eq:rd_theory}), 
respectively. As a result, the high-to-low ratio decreases as compared with the traditional Glauber calculations in the GEA framework~\cite{Sargsian97}. 

 As shown in Figure~\ref{fig:rescatteringPrediction}, an increasing $Q^2$-dependence of the FSIs when compared to ``no CT'' predictions could indicate a regime for the onset of CT.  This allows us to experimentally explore proton CT in kinematics that was elusive or inaccessible in previous $(e,e'p)$ experiments. Furthermore, comparing cross-section ratios between data and theory relies less on the model dependence than comparing the absolute cross section to calculations directly since some assumptions in the theory are canceled in the ratio.

\section{A Possible Near-Term Hall C Experiment}
\label{sec:new_proposal}
With near-term-planned upgrades to the JLab beamline, it will be possible for Hall C to receive an incident 11~GeV electron beam energy. Hall C is uniquely capable of measuring high luminosity and high precision and has already been used in the a $^{12}$C$(e,e'p)$ color transparency experiment~\cite{PhysRevLett.126.082301} and a deuterium electro-disintegration experiment~\cite{cyero_2020}. This 11~GeV beam energy, together with the full momentum capabilities of the Hall C spectrometers, can extend the measurable $Q^2$ up to 17~(GeV$/c)^2$. Here, we discuss the possibility to measure the onset of CT in D$(e,e'p)n$ by detecting the electron and proton in coincidence in Hall C from electron scattering on a liquid deuterium target. Hall C is comprised of two spectrometer arms: the high momentum spectrometer (HMS) and the super high momentum spectrometer (SHMS)~\cite{hms_info,shms_nim}. The spectrometer pair is capable of detecting particles with momenta up to 7~GeV/$c$ with a precision on the order of 0.1\% (HMS), and 11~GeV/$c$ with a precision on the order of 0.05\% (SHMS). Note that the spectrometer momentum and angular range, not the beam energy, set the limit on our $Q^2$ coverage. Both spectrometers are equipped with standard particle detectors, including time-of-flight scintillators for fast timing and triggering, drift chambers for precise particle tracking, and Cherenkov detectors and electromagnetic calorimeters for detailed particle identification. 

In such an experiment, for each $Q^2$ setting, the HMS would detect the scattered electrons at the quasi-elastic peak, while the struck protons would be detected in the SHMS with high momentum resolution ($\sim$10 MeV/$c$). The SHMS momentum acceptance could be centered, where $p_r=200$ MeV$/c$, which would enable coverage for both the low (\mbox{50--150 MeV$/c$}) and high (300--600 MeV$/c$) $p_r$ regions simultaneously due to its large momentum acceptance. The experimental observable is the ratio of the measured $(e,e'p)$ cross sections from those two momentum regions:
\begin{equation}
R_D^{\text{exp}}(Q^2)=\dfrac{\sigma(300\leq p_r \leq 600 ~{\rm MeV/}c; Q^2)\downarrow}{\sigma(50\leq p_r\leq 150~{\rm MeV/}c; Q^2)\uparrow} \,.
\end{equation}

 {As shown in Figure~\ref{fig:rescatteringPrediction}, $R^{\text{exp}}_D$ is expected to decrease with increasing $Q^2$ in the presence of PLCs. 
We also require the neutron recoil angle $\theta_{nq}$ of 60$^{\circ}$
to 90$^{\circ}$ to maximize FSI effects in both $p_r$ regions so that the expected change in the ratio is maximized in the presence of CT. Note that the FSI effect is more sensitive to the choice of wave functions at high $p_r$, e.g., estimates with CD-Bonn instead of AV18 give less than $5\%$ more FSI amplitude at $p_r=400$ MeV but the discrepancy increases to 20\% at $p_r=600$ MeV. However, this is not a significant contribution to the integrated cross section at the high $p_r$ (300--600 MeV$/c$) region due to the rapid falling of cross sections at large $p_r$. Furthermore, this discrepancy can be suppressed if one normalizes the experimental ratios to the same ratios calculated with PWIA. A detailed study can be conducted in the future to identify the best value of $p_r$ optimized for maximal FSI effect and minimal sensitivity with respect to the choice of the deuteron wave function.}

\begin{figure}[H]%[htb!]
%\centering
\includegraphics[width=0.85\textwidth]{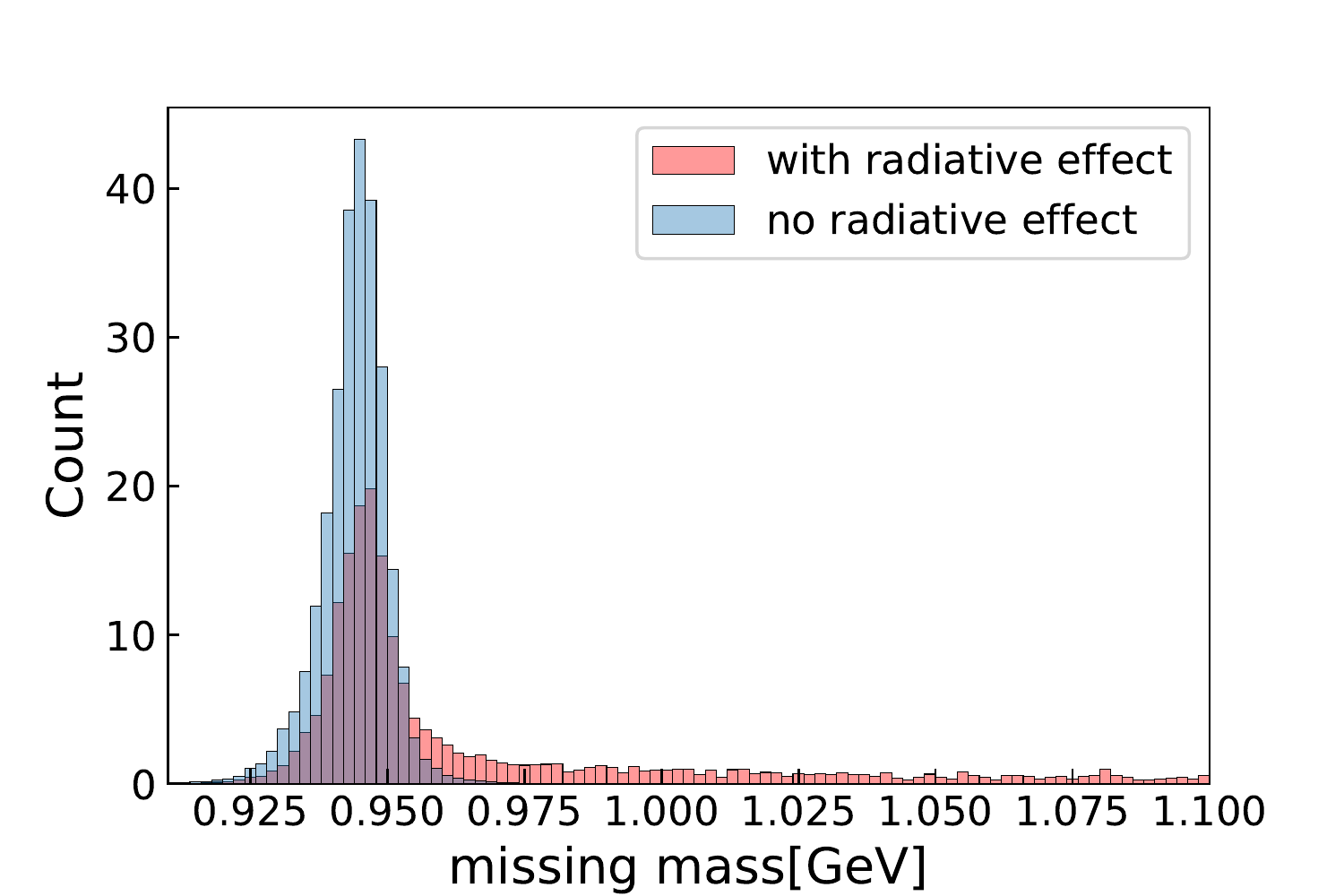}
\caption{\label{fig:missing_mass} The missing mass from the Hall C simulation software %, SIMC,
(SIMC Monte Carlo package~\cite{simc,Dutta:2000sn})  
with (red) and without (blue) radiative effects at $Q^2=14$~(GeV$/c)^2$.}
\end{figure}

The missing mass is an essential quantity to reconstruct for all measured kinematics. Kinematics and rate studies performed with the Hall C 
SIMC Monte-Carlo simulation package~\cite{simc,Dutta:2000sn} show the reconstructed missing mass of the neutron with and without radiative effects 
as shown in Figure~\ref{fig:missing_mass}. The estimated detectable rates rely on a cut on the reconstructed neutron mass. 

 {With a $25$~cm liquid deuterium target (a total of 5$\%$ radiation lengths) and a 80~$\upmu$A continuous electron beam, this proposed experiment can reach a luminosity per proton of 6.23 $\times 10^{38}$~A$^{-1}$cm$^{-2}$s$^{-1}$. Rate estimations with SIMC and the AV18+GEA cross-section model as discussed above show that, with three months of beam time, we can determine the cross-section ratio $R_D$ at $Q^2=8, 10, 12, 14, 15$ (GeV$/c)^2$ within statistical $3\%$ uncertainties (i.e., at least $5$ sigma sensitivities at $\delta M^2=3$ GeV$^2$). The $8$ to $14$ (GeV$/c)^2$ settings overlap with the previous $^{12}$C$(e,e'p)$ experiment that searched for the onset of CT signals in parallel kinematics. The $15$ (GeV$/c)^2$ setting, which takes two months of the estimated beam time, would push the proton CT search via $(e,e'p)$ to a record high $Q^2$ to improve the size of signal, and test new theoretical predictions\cite{Brodsky:2022bum}. All of this experimental design is modeled under reasonable Jefferson Lab capabilities. The central momenta ranges for the SHMS and HMS spectrometers range from 5.1 to 8.9~GeV/$c$ (SHMS) and 2.9 to 6.7~GeV/$c$ (HMS), well within the current operating ranges of each spectrometer. The maximum $Q^2$ of 15~(GeV$/c)^2$) is chosen based on the rates and reasonable running time with the electron beam.}

While a higher beam energy would not increase the $Q^2$ coverage due to the spectrometer limitations, increasing the beam energy by 0.5~GeV would increase the rates by $\sim$30\% and could significantly reduce beam time and/or improve statistical precision. 
 {The systematic uncertainty of such an experiment would be comparable to that of the recent color transparency experiment~\cite{PhysRevLett.126.082301}. In that experiment, the total systematic uncertainty was determined to be 4\% with the largest contribution (2.6\%) due to the uncertainty in the spectrometer acceptance. The next largest sources of uncertainty were the knowledge of the fundamental $ep$ elastic cross section (1.8\%) and the proton absorption in materials between the target and detector (1.2\%). Both of these effects will be minimized as systematic contributions through the use of experimental ratios. }

\section{Summary}
We propose a realistic near-term experimental effort to search for the onset signal of CT using a deuterium target at JLab that could be accomplished with approximately 3 months of 11~GeV electron beam time. The experiment would probe a $Q^2$ range from 8--15~GeV$^2$ and would have an increased sensitivity as compared to previous electron-scattering experiments, allowing it to better observe or rule out a signal for the onset of CT in the proton for this momentum range. The deuteron is advantageous for such a measurement because the CT onset signal is expected to be enhanced by measuring knockout reactions in kinematical regions already dominated by high FSIs and is described well in terms of the generalized eikonal approximation at high momentum. Double-scattering effects have a unique enhancement to the cross section ratio that is well-studied experimentally and has a clear dependence on the recoiling neutron angle. JLab Hall C is the first choice to perform such a measurement, but other experimental setups could be explored with upgrades to detector acceptances, luminosities and electron beam energy. {The only needed improvements from the current Hall C operations for this experiment would be the accelerator's capability to deliver an 11~GeV energy electron beam (currently operating about 4\% lower than this) which is planned for in near-term accelerator upgrades. }

\vspace{6pt} 
\authorcontributions{
All authors have contributed equally to this manuscript and have read and agreed to the published version of the manuscript.}

\funding{This research was funded in part by Department of Energy (DOE) grant number DE-AC05-06OR23177 
under which the Jefferson Science Associates operates the Thomas Jefferson National Accelerator Facility. 
S.L. 
is supported by Department of Energy, Office of Nuclear Physics, under contract number DE-AC02-05CH11231. 
C.Y.  
is supported by the National Science Foundation MPS-Ascend Postdoctoral Research Fellowship under award number 2137604. 
J.R.W.
is supported by the Laboratory Directed Research and Development (LDRD)
programs of the Lawrence Berkeley National Laboratory (LBNL)
the Electron-Ion Collider (EIC)
Center at Jefferson Lab and by the U.S. Department of Energy, Office of Science, Office of Nuclear Physics, under contract number DE-AC02-05CH11231.  
W.C. is supported by the National Science Foundation under Award No. 2111442. 
C.B. 
acknowledges support by Department of Energy, Office of Nuclear Physics under contract DE-SC0022007.  
M.S.'s 
work is supported by the U.S. DOE Office of Nuclear Physics grant DE-FG02-01ER41172.}

\dataavailability{Not applicable. 
} 

\conflictsofinterest{The authors declare no conflict of interest.}

%%%%%%%%%%%%%%%%%%%%%%%%%%%%%%%%%%%%%%%%%%
\begin{adjustwidth}{-\extralength}{0cm}

\reftitle{References}

%\end{paracol}
%=====================================
% References
%=====================================
%\bibliographystyle{ieeetr}
%\bibliography{refs.bib}

\end{adjustwidth}
\end{document}